\documentclass[11pt]{article}

\usepackage{amsmath,amssymb,amsfonts,bbm,bm}
\usepackage{slashed}

\usepackage{tocloft} 

\usepackage[nosort]{cite} 
\usepackage{graphicx,color}
\usepackage[colorlinks=true, linkcolor=blue, citecolor=blue, linktoc=all]{hyperref}

\usepackage[usenames,dvipsnames]{xcolor}

\usepackage{tikz-cd}
\usepackage{pict2e}
\usepackage{physics}
\usepackage{comment} 
\usepackage[makeroom]{cancel}

\newcommand{\beq}{\begin{eqnarray}}
\newcommand{\eeq}{\end{eqnarray}}
\newcommand{\beqn}{\begin{eqnarray}}
\newcommand{\eeqn}{\end{eqnarray}}
\newcommand{\bea}{\begin{eqnarray}}
\newcommand{\eea}{\end{eqnarray}}
\newcommand{\be}{\begin{equation}}
\newcommand{\ee}{\end{equation}}

\newcommand{\un}[1]{\underline{#1}}

\usepackage[normalem]{ulem}

\newcommand{\caltech}[1]{
  \centerline{
    \begin{minipage}[c]{0.7\textwidth}
      \begin{center}
      ${}^{#1}$ Walter Burke Institute for Theoretical Physics,\\
      California Institute of Technology, Pasadena, CA 91125, U.S.A.
      \end{center}
    \end{minipage}
  }
}

\usepackage{mathrsfs,mathtools}
\renewcommand\mathbb[1]{\mathbbm{#1}}

\makeatletter
\DeclareRobustCommand{\loplus}{\mathbin{\mathpalette\dog@lsemi{+}}}
\DeclareRobustCommand{\lotimes}{\mathbin{\mathpalette\dog@lsemi{\times}}}
\DeclareRobustCommand{\roplus}{\mathbin{\mathpalette\dog@rsemi{+}}}
\DeclareRobustCommand{\rotimes}{\mathbin{\mathpalette\dog@rsemi{\times}}}

\newcommand{\dog@rsemi}[2]{\dog@semi{#1}{#2}{-90,90}}
\newcommand{\dog@lsemi}[2]{\dog@semi{#1}{#2}{270,90}}
\newcommand{\dog@semi}[3]{%
  \begingroup
  \sbox\z@{$\m@th#1#2$}%
  \setlength{\unitlength}{\dimexpr\ht\z@+\dp\z@\relax}%
  \makebox[\wd\z@]{\raisebox{-\dp\z@}{%
    \begin{picture}(1,1)
    \linethickness{\variable@rule{#1}}
    \roundcap
    \put(0.5,0.5){\makebox(0,0){\raisebox{\dp\z@}{$\m@th#1#2$}}}
    \put(0.5,0.5){\arc[#3]{0.5}}
    \end{picture}%
  }}%
  \endgroup
}
\newcommand{\variable@rule}[1]{%
  \fontdimen8  
  \ifx#1\displaystyle\textfont3\else
    \ifx#1\textstyle\textfont3\else
      \ifx#1\scriptstyle\scriptfont3\else
        \scriptscriptfont3\relax
  \fi\fi\fi
}
\DeclareRobustCommand{\loplus}{\mathbin{\mathpalette\dog@lsemi{+}}}

\usepackage{amsthm}

\renewcommand{\op}[1]{\boldsymbol{#1}}

\newcommand{\thistitle}{A Theory of Backgrounds and Background Independence}

\begin{document}

\title{\thistitle}
\author{
	Marc S. Klinger 
	\\
	\\
	{\small \emph{\caltech{}}}
	\\
	}
\date{}
\maketitle
\vspace{-0.5cm}
\begin{abstract}
\vspace{0.3cm}
In this note, we describe how the study of backgrounds for general quantum systems can be formulated in terms of the representation theory of abstract $C^*$ algebras. We illustrate our general framework through two example systems: superconductivity and perturbative quantum gravity. In both cases, spontaneously broken symmetries imply the existence of unitarily inequivalent Hilbert spaces that play the role of distinct backgrounds relative to which observables are measured. Background independence can be realized by gauging the broken symmetry; extending the algebra of observables for the theory to include new physical processes that intertwine between these disjoint representations. From the point of view of the background independent theory, different backgrounds have an interpretation as different vacuum expectation values of these intertwining operators. In superconductivity, the intertwiners are intimately related to the Josephson effect. In gravity, they are related to geometric fluctuations. We explain how this framework is connected to recent work on generalized symmetries and algebraic extensions. 
To this end, we close with some remarks about how the operator algebra of a closed universe may arise from a generalized symmetry associated with an algebraic extension of the causal wedge by appealing to subregion-subalgebra duality. 
\end{abstract}

\begingroup
\hypersetup{linkcolor=black}
\tableofcontents
\endgroup

\setcounter{footnote}{0}
\renewcommand{\thefootnote}{\arabic{footnote}}

\newcommand{\curr}[1]{\mathbb{J}_{#1}}
\newcommand{\constr}[1]{\mathbb{M}_{#1}}
\newcommand{\chgdens}[1]{\mathbb{Q}_{#1}}
\newcommand{\spac}[1]{S_{#1}}
\newcommand{\hyper}[1]{\Sigma_{#1}}
\newcommand{\chg}[1]{\mathbb{H}_{#1}}
\newcommand{\ThomSig}[1]{\hat{\Sigma}_{#1}}
\newcommand{\ThomS}[1]{\hat{S}_{#1}}
\newcommand{\discuss}[1]{{\color{red} #1}}
\theoremstyle{definition}
\newtheorem{theorem}{Theorem}[section]
\newtheorem{example}{Example}[section]
\newtheorem{definition}{Definition}[section]

\pagebreak

\section{Introduction}

The goal of this note is to shed new light on recent work concerning the algebraic approach to quantum field theory, gauge theory, and quantum gravity \cite{Klinger:2023tgi,Klinger:2023auu,AliAhmad:2024wja,AliAhmad:2024eun,AliAhmad:2024vdw,AliAhmad:2025oli,Klinger:2025hjp,AliAhmad:2025bnd,KKFS:2025,Witten:2021unn,Chandrasekaran:2022cip,Chandrasekaran:2022eqq,Witten:2023xze,Kudler-Flam:2023qfl,Chen:2024rpx,Jensen:2023yxy,Faulkner:2024gst,AliAhmad:2023etg}. Rather than dwell too much on the details of what is indeed a very technical subject, our intention is to illuminate one rather appealing aspect of this program through a series of two examples: superconductivity and perturbative quantum gravity. The idea we are most interested in presenting is how the enlarged perspective on algebras of observables adopted in recent work provides a new lens through which to study infrared phenomena, especially in relation to vacuum physics and superselection. Understanding the vacuum is a crucial step towards contending with the problem of background independence, and we propose a novel framework to this end. 

A very important difference between the approach of the present paper and other related work is the focus on $C^*$ algebras of observables in contrast to von Neumann algebras.\footnote{This distinction has been emphasized in \cite{AliAhmad:2024wja,AliAhmad:2024vdw,AliAhmad:2025bnd,Witten:2023xze,Liu:2025cml,Liu:2025krl}.} More to the point, we concentrate largely on \emph{abstract} $C^*$ algebras which are defined independently of a preferred Hilbert space representation. From this point of view, different `vacua' correspond to different representations \cite{Haag:1992hx,Haag:1962msc}. This correspondence can be drawn rather explicitly by appealing to the Gelfand-Naimark-Segal (GNS) construction in which a representation of an abstract $C^*$ algebra $A$ is derived from a chosen state on this algebra \cite{gelfand1943imbedding,segal1947irreducible}.\footnote{See Appendix \ref{app: GNS} for an introduction to the GNS construction.} Here, by state, we simply mean a positive, linear, normalized function from $A$ to the complex numbers which can be interpreted as an expectation value. A pair of states, $\varphi,\psi$, on a common algebra are said to be `quasi-equivalent' if the share the same set of operators with zero-expectation values. If instead these states assign zero expectation values to different sets of operators their GNS representations are unitarily inequivalent. In the parlance of quantum mechanics, the states are superselected. There can be no allowed observable inside the algebra $A$ which `takes the state $\varphi$ to the state $\psi$'. 

To interpret the latter statement, it is useful to pass to what is called the universal representation of the algebra $A$ \cite{Kadison:Ringrose1991}. Let $S_0(A)$ denote the quotient of the complete set of states on $A$ by the equivalence relation described above. Then, the universal representation of $A$ is a direct sum of the GNS representations of $\varphi \in S_0(A)$. Schematically
\beq
	\mathscr{H}_{U}(A) \equiv \bigoplus_{\varphi \in S_0(A)} \mathscr{H}_{\varphi}, 
\eeq
and we denote the associated representation by $\pi_{U}: A \rightarrow B(\mathscr{H}_{U}(A))$. Each state $\varphi \in S_0(A)$ has a vector representative $\ket{\varphi} \in \mathscr{H}_{U}(A)$ such that $\varphi(a) = \bra{\varphi} \pi_{U}(a) \ket{\varphi}$. If $\varphi$ and $\psi$ are inequivalent states there will exist no $a \in A$ such that $\bra{\varphi} \pi_{U}(a) \ket{\psi} \neq 0$.  

In quantum field theory, the choice of vacuum state is intimately tied to the choice of background spacetime \cite{Witten:2023xze,Liu:2025krl,Afshordi:2012jf,Brum:2013bia,Hawkins:2022hud}. The abstract $C^*$ algebra of interest is roughly the canonical commutation/anticommutation relation (CCR/CAR) algebra of a free field theory which we denote by $A$.\footnote{At this stage we are still treating gravity as non-dynamical within the algebra $A$. Other interactions can be treated perturbatively, see e.g. \cite{DAngelo:2022vsh}.} The typical approach to specifying this algebra requires selecting a preferred spacetime from the onset, since the algebraic relations depend on the metric through the classical equations of motion and its propagators. However, a very elegant approach to circumventing this problem has been proposed by Brunetti, Fredenhagen, and Rejzner going under the name of locally covariant quantum field theory \cite{Brunetti:2006qj,Fredenhagen:2011hm,Brunetti:2013maa,Brunetti:2022itx}. The idea is to `construct the theory simultaneously on all spacetimes in a coherent way' \cite{Fredenhagen:2011hm}. Consequently, the algebra $A$ is operationally well defined regardless of what spacetime we choose.\footnote{Here and henceforward by spacetime we really mean a globally hyperbolic spacetime.} This idea seems closely related to recent work by Witten in which he argued that a background independent algebra should be defined in terms of universal data that makes sense in any spacetime \cite{Witten:2023xze}. The importance of the background then only becomes pronounced when we choose a particular Hilbert space representation in which allowed states correspond to positive energy excitations in some geometry. 

Moving into the universal representation of the algebra $A$ we see that states representing different geometries are superselected. This is simply a consequence of the fact that the algebra $A$ does not include operations corresponding to geometric fluctuations. From the point of view of quantum field theory the geometry is, by definition, a background feature. 
All measurements and expectation values in the theory are referred to this fixed background. Naturally, this cannot be the case in quantum gravity in which geometry changing operations should be commonplace. In this sense, it seems that the best way to overcome the problem of background independence is to incorporate such processes while maintaining quantum field theory as a fixed geometry limit. 

We will not be implementing the above recommendation for fully non-perturbative quantum gravity in this paper. Instead, we will explore two simpler examples in which the same problems/structure arise and can be dealt with. The first example, described in Section \ref{sec: BCS}, is the Bardeen-Cooper-Schrieffer (BCS) theory of superconductivity \cite{Bardeen:1957mv,Bardeen:1957kj}. BCS famously exhibits a degenerate vacuum which is associated with the spontaneous breaking of a $U(1)$ global symmetry. As we will discuss, spontaneously broken symmetries rather generically coincide with situations in which some physical processes are `hidden' behind the choice of a background. In the case of superconductivity, these hidden processes are those which change the number of Cooper pairs in the superconducting sample or the overall phase of its many-body quantum state. 

To make these processes `visible' we must introduce new interactions and/or constraints. For the BCS theory this entails bringing the superconductor into contact with a bath. For simplicity, we will conceptualize this bath as a second superconducting system separated from the first by an insulating barrier called a Josephson junction. Cooper pairs can now be tunneled from one superconductor to the other resulting in a measurable Josephson current \cite{Josephson:1962zz}. This current is intimately related to transitions between would-be vacuum states for the isolated superconducting system. As a result, the algebra of observables of either superconducting subsystem includes a new operator that measures the relative phase between the superconducting states (or the relative number of Cooper pairs). Mathematically, this enlarged algebra is of the form of a $C^*$ crossed product between the naive superconducting algebra and the spontaneously broken $U(1)$ symmetry group. 

Our second example, described in Section \ref{sec: Gravity}, is perturbative quantum gravity around a background spacetime admitting a bifurcate Killing horizon. We concentrate specifically on the region of this spacetime, $\mathscr{R}$, which is accessible to an infinitely long-lived observer. A von Neumann algebraic analysis of the same setting has been investigated in \cite{KKFS:2025}, with the primary conclusions concerning the study of the entropy of states in the resulting gravitational algebra. In that work, only the unbroken gravitational symmetries were addressed to ensure that the resulting gravitational algebra retained a consistent interpretation at leading order in perturbation theory. In this paper we will discuss how the broken symmetries, which can implement constraints at higher perturbative order, may be gauged and provide a physical interpretation of the geometric operators realized in the resulting algebra. Rather than follow the same approach as the BCS example, in this example we simply quantize the theory as an abstract $*$-algebra and then explore its relation to vacuum physics by constructing an appropriate representation. We find that, on account of gravitational constraints which entangle the region $\mathscr{R}$ and its complement, new operators must be appended to the algebra of the subregion which encode geometric fluctuations of the background spacetime. The resulting Hilbert space representation has the form of a direct integral over GNS representations for the linear graviton theory with respect to states associated with the various deformed geometries. The algebra of the subregion includes operators that create coherent excitations over a fixed background, and gravitational charges that intertwine backgrounds. 

These charges can be viewed explicitly as representations of superboosts and superrotations that act on the system in such a way as to preserve the subregion $\mathscr{R}$. This is a generalization of the perhaps more familiar example of the BMS group which preserves the class of asymptotically flat spacetimes \cite{Sachs:1962zza}. The algebraic description of the subregion in light of the gravitational constraints therefore resolves at least \emph{some} of the background independence problem implicit to quantum field theory. We can now move freely between (QFT) vacua corresponding to different background metrics. However, we are only allowed to move within the class of metrics that are diffeomorphic to some `overall background' by a superboost or superrotation. This limitation seems quite closely related to the observation made in \cite{Liu:2025cml,Liu:2025krl} that \emph{concrete} operator algebras obtained from a quantum gravity theory, like string theory, are implicitly referenced to a class of asymptotic geometries. Presumably, a complete description of quantum gravity should also involve intertwining operations between these `overall backgrounds'. 

Moving toward this goal, in Section \ref{sec: Discussion}, we provide a more abstract summary of our framework. We first emphasize that new observables may be accessible even from the perturbative gravitational algebra, especially through the analogy to the Josephson effect. We then make some speculative remarks about how the framework described here may help address the problem of background independence in fully non-perturbative quantum gravity. These comments build upon recent work concerning the relationship between the representation theory of a $C^*$ algebra, its generalized symmetries \cite{AliAhmad:2025bnd}, and its possible extensions into larger algebras \cite{AliAhmad:2025oli}. In subsection \ref{App: BU}, we describe an application of this point of view to the problem of holography for closed universes. Building upon the recent work \cite{Liu:2025cml}, we argue that the algebra of baby universe operators can be realized as a quantum symmetry which is broken in effective field theory but can be repaired by a generalized form of gauging. This point of view provides a purely algebraic avatar for entanglement wedge reconstruction which builds upon a related proposal made in \cite{AliAhmad:2024saq}.

\section{Lessons from Superconductivity} \label{sec: BCS}

Superconductivity provides a useful case study through which to illustrate our notion of backgrounds and background independence. We follow Haag's algebraic approach to the BCS model \cite{Haag:1962msc}. Our presentation will be rather sparse, as we are most interested in superconductivity as an example of a general set of ideas rather than for its specific features. 

\subsection{Basic set up}

The relevant algebra for describing the BCS theory is the canonical anticommutation relation (CAR) algebra which we denote simply by $A$. This algebra is generated by spinor fields on a spatial surface $\Sigma$, $\overline{\psi}_r(x)$ and $\psi^s(x)$, satisfying the anticommutation relation
\beq
	\{\overline{\psi}_r(x), \psi^s(y)\} = i\delta^s_r \delta(x-y). 
\eeq
Rigorously, we regard $\overline{\psi}_s(x)$ and $\psi^r(x)$ as operator valued distributions e.g. so that the relevant operators are smeared fields
\beq
	\overline{\psi}[f] \equiv \int_{\Sigma} d\mu(x) f^s(x) \overline{\psi}_s(x), \qquad \psi[g] \equiv \int_{\Sigma} d\mu(x) g_r(x) \psi^r(x). 
\eeq
The physical description of the BCS theory is completed by specifying its dynamics. In addition to the usual quadratic piece, the BCS Hamiltonian contains a four point interaction\footnote{Here, we have introduced the notation $d\mu^{\otimes n}(x_1, ..., x_n) = d\mu(x_1) ... d\mu(x_n)$. 
}
\beq
	U = \int_{\Sigma^4} d\mu^{\otimes 4}(x,x',y,y') 
	\; v(y,y') \; \overline{\psi}_1(x) \overline{\psi}_2(x+y) \psi^2(x'+y') \psi^1(x'). 
\eeq

Although the BCS Hamiltonian is not free, it can be diagonalized exactly by exploiting properties of the canonical anticommutation relations. A typical quasi-local observable is of the form
\beq
	Q(x) \equiv \int_{\Sigma^{2n}} d\mu^{\otimes 2n}(y_1, ..., y_n, z_1, ..., z_n) F(y_i, z_j)^{r_1 ... r_n}_{s_1 ... s_n} \prod_{i = 1}^n \overline{\psi}_{r_i}(x+y_i) \prod_{j = 1}^n \psi^{s_j}(x + z_j). 
\eeq
The spatial average of such an observable, 
\beq
	\overline{Q} \equiv \lim_{V \rightarrow \infty} \frac{1}{V} \int_{\Sigma} d\mu(x) Q(x),
\eeq	
can be seen to commute with all smeared field operators
\beq
	\bigg[\overline{Q},\psi[g]\bigg] = \bigg[\overline{Q},\overline{\psi}[f]\bigg] = 0. 
\eeq
As such, if $\pi: A \rightarrow B(\mathscr{H})$ is an irreducible representation\footnote{A representation $\pi: A \rightarrow B(\mathscr{H})$ is called irreducible if there are no bounded operators on $\mathscr{H}$ which do not belong to the weak closure of $\pi(A)$. In other words, $\pi(A)'' = B(\mathscr{H})$. There may be some confusion as to why we are allowed to consider an irreducible representation in this context. Typically, one says that the algebra of observables for a quantum field theory is of type III and therefore admits no irreducible representations. In fact, this statement applies to algebras of quantum fields restricted to a spacetime subregion. The fact that such algebras admit no irreducible representations is tantamount to the statement that they are `infinitely' entangled with their complementary regions. For our purposes, we are interested in teh global algebra of quantum fields whose weak closure is generically of type I.} of $A$ then $\pi(\overline{Q}) \in Z(\pi(A)) \simeq \mathbb{C}$ is a complex number. 

While the interaction potential $U$ does not have the form of a quasi-local observable, the above allows us to conclude that the operator
\beq
	W = \int_{\Sigma^2} d\mu^{\otimes 2}(x,y) \bigg( \Delta(y) \overline{\psi}_1(x) \overline{\psi}_2(x+y) + \overline{\Delta(y)} \psi^2(x+y) \psi^1(x) \bigg)
\eeq	
can replace $U$ in the BCS Hamiltonian provided we are working in an irreducible representation. That is $\pi(U)$ and $\pi(W)$ have the same algebraic relations with observables in $\pi(A)$. The function
\beq \label{Gap Eqn}
	\Delta(y) \equiv \int_{\Sigma} d\mu(z) v(y,z) \varphi(z)
\eeq	
is expressed in terms of the coupling $v(y,z)$ and a to-be-determined function $\varphi(z)$. Eqn. \eqref{Gap Eqn} is called the gap equation. 

The BCS Hamiltonian is therefore quadratic in the fields within a given irreducible representation. As such, it can be diagonalized by performing an appropriate Bogoliubov transformation. The unknown function $\varphi$ appearing in the gap equation can be regarded as determining the overall vacuum state of the BCS theory. In particular,
\beq
	\varphi(z) \equiv \omega_{\varphi}( \psi^2(z) \psi^1(0) )
\eeq
is the Feynman propagator of the spinor field which defines a quasi-free state on the algebra $A$. The irreducible representation associated with a given $\varphi$ is identified with the GNS representation of the state $\omega_{\varphi}$, $\pi_{\omega_{\varphi}}: A \rightarrow B(\mathscr{H}_{\varphi})$. Within this representation, the spatial average of a quasi-local operator, which was shown to be a $\mathbb{C}$-number, becomes equivalent to the expectation value
\beq
	\pi_{\varphi}(\overline{Q}) = \omega_{\varphi}(Q(x)). 
\eeq
There exist a one-parameter family of solutions to the gap equation $\varphi_{\phi}$ with $\phi \in U(1)$. Thus, we conclude that the BCS Hamiltonian admits a $U(1)$ family of degenerate vacua which we label by $\omega_{\varphi_{\phi}} \equiv \omega_{\phi}$. 

\subsection{Symmetry Breaking: Degenerate Vacua}

It was Haag who first recognized that the appearance of these degenerate vacua can be traced back to a spontaneously broken symmetry. In the algebraic formalism a symmetry is a map $\alpha: A \rightarrow A$ which preserves the algebraic relations on $A$ e.g.
\beq
	\alpha(ab) = \alpha(a) \alpha(b), \qquad \alpha(a^*) = \alpha(a)^*, \qquad \norm{\alpha(a)} = \norm{a}. 
\eeq
The collection of such maps forms a group called the automorphism group of $A$ which we denote by $\text{Aut}(A)$. Let $\pi: A \rightarrow B(\mathscr{H})$ be a Hilbert space representation. The symmetry $\alpha \in \text{Aut}(A)$ is spontaneously broken in the representation $\pi$ if there does not exist a unitary operator $U \in U(\mathscr{H})$ which implements $\alpha$ e.g. such that
\beq
	\pi \circ \alpha(x) = U \pi(x) U^{\dagger}. 
\eeq 

The algebraic notion of spontaneous symmetry breaking becomes most clearly related to the more conventional notion by passing through the GNS construction. Given a state $\omega \in S(A)$ and a symmetry $\alpha \in \text{Aut}(A)$ we can define a new state $\omega_{\alpha} \equiv \omega \circ \alpha$. If $\alpha$ is spontaneously broken in the GNS representation $\pi_{\omega}$, then the representations $\pi_{\omega}$ and $\pi_{\omega_{\alpha}}$ are unitarily inequivalent. As such, we conclude that there exist no unitarily implementably processes which map the state $\omega$ to the state $\omega \circ \alpha$. In more standard parlance, the states $\omega$ and $\omega \circ \alpha$ are \emph{superselected}. 

Returning to the case of superconductivity, the CAR algebra $A$ admits a $U(1)$ global symmetry $\alpha: U(1) \rightarrow \text{Aut}(A)$
\beq
	\alpha_{\phi}(\psi[g]) = e^{i \phi} \psi[g], \qquad \alpha_{\phi}(\overline{\psi}[f]) = e^{-i\phi} \overline{\psi}[f]. 
\eeq
However, this symmetry is spontaneously broken with respect to any of the vacuum representations $\pi_{\phi} \equiv \pi_{\varphi_{\phi}}$, which also coincide with the GNS representations of the vacua $\omega_{\phi}$. In particular, 
\beq
	\pi_{\phi} \circ \alpha_{\phi'} = \pi_{\phi + 2\phi'}. 
\eeq	
Thus, we conclude that there are no unitarily implementable processes accessible within the algebra $A$ that transform a given vacuum state $\omega_{\phi}$ into a different vacuum state $\omega_{\phi'}$. 

A useful way to understand a spontaneously broken symmetry is that it `wants' to be implemented by an operator that is `unobservable'. This somewhat imprecise heuristic is made clear in the BCS example. The spontaneously broken $U(1)$ symmetry may be associated either with the operator $N$ which counts the overall number of Cooper pairs within the superconductor or the operator $\Phi$ which measures the phase of the many-body quantum state of the Cooper pairs. For a single isolated superconductor, neither of these quantities constitute good observables.

This point of view is best established by considering an enlarged Hilbert space in which the operators $N$ and $\Phi$ are well defined but remain unobservable. In particular, we can consider the direct integral Hilbert space $\mathscr{H}_{ext} \equiv \int^{\oplus}_{U(1)} d\phi \; \mathscr{H}_{\phi}$. A nice way to describe states in this Hilbert space is to consider an `overall vacuum'
\beq
	\ket{\Omega} \equiv \int_{U(1)}^{\oplus} d\phi \ket{\Omega_{\phi}},
\eeq	
where $\ket{\Omega_{\phi}}$ is the vector representative of the vacuum state $\omega_{\phi}$ within its GNS representation. The most general operator on $\mathscr{H}_{ext}$ is of the form
\beq \label{General op on ext}
	\mathfrak{X} \equiv \int_{U(1)}^{\oplus} d\phi \; \pi_{\phi}(\mathfrak{X}(\phi)),
\eeq
where $\mathfrak{X}: U(1) \rightarrow A$. This operator acts on the overall vacuum to create a state
\beq
	\ket{\mathfrak{X}} \equiv \mathfrak{X} \ket{\Omega} = \int_{U(1)}^{\oplus} d\phi \; \pi_{\phi}(\mathfrak{X}(\phi)) \ket{\Omega_{\phi}} = \int_{U(1)}^{\oplus} d\phi \ket{\mathfrak{X}(\phi)}_{\phi}.
\eeq

On the Hilbert space $\mathscr{H}_{ext}$ we have access to a pair of new operators which make the symmetry breaking picture more clear. Firstly, there is an operator $\Phi$ which acts as a `position operator' relative to the vacuum phase
\beq
	\Phi \ket{\Omega_{\phi}} \equiv \phi \ket{\Omega_{\phi}}.
\eeq	
In this regard, we can view the eigenvalues of the phase operator in the extended Hilbert space as labeling the degenerate vacua, or, equivalently, the superselection sectors of the BCS theory. The conjugate operator $N$ acts as a `momentum operator' relative to the phase generating translations\footnote{The factor of two comes from the fact that the fundamental quanta being counted by $N$ is a Cooper pair.}
\beq
	e^{i\phi N} \ket{\Omega_{\phi'}} \equiv \ket{\Omega_{\phi' + 2\phi}}.
\eeq
We can also define a representation of $A$ on $\mathscr{H}_{ext}$ by
\beq
	\pi(a) \ket{\Omega_{\phi}} \equiv \pi_{\phi}(a) \ket{\Omega_{\phi}}. 
\eeq
Then, it is straightforward to show that
\beq
	e^{i \phi N} \pi(a) e^{-i\phi N} = \pi \circ \alpha_{\phi}(x),
\eeq
meaning $e^{i\phi N}$ is precisely the one-parameter unitary group that would like to generate the automorphism $\alpha$. However, $e^{i \phi N}$ do not belong to the algebra $A$ and thus remain unobservable.  

From the perspective of the enlarged Hilbert space $\mathscr{H}_{ext}$ the choice of a vacuum state can be likened to a choice of non-dynamical quantum reference frame \cite{AliAhmad:2024wja,AliAhmad:2024vdw,Bartlett2007}, or, more suggestively, a background. The fact that $\Phi$ and $N$ are not contained inside the algebra $A$ means that their fluctuations are inaccessible to an experimenter who observes only the algebra $A$. No physically observable processes can change the value of $\phi$ once it is chosen and thus $\phi$ constitutes a background or reference value relative to which all subsequent measurements are made. 

\subsection{Symmetry Repairing: Vacuum Transitions}

In the context of superconductivity something interesting happens which is that each $\pi_{\phi}(A)$ and $\pi_{\phi'}(A)$ for $\phi \neq \phi'$, while unitarily inequivalent, are in some sense `the same'. This is because there exists a subclass of observables, those which are invariant under the $U(1)$ broken symmetry, for which the expectation values computed in any representation are the same. With this being said, we have already seen that for other observables -- like the spatial average of a quasi-local quantity -- expectation values \emph{do} depend on the choice of background. This demands the question, how can we promote our theory to be `background independent'?

One way to achieve `background independence' is to `gauge' the broken symmetry. Here, we are using the term gauging in the path integral sense e.g. when a background field is promoted to a dynamical one. Physically, the gauging of a broken symmetry involves introducing new interactions and constraints which promote the previously unobservable operators that define the broken symmetry to the status of observable. Once again, superconductivity provides us with a stark example of this phenomenon through the Josephson effect \cite{Josephson:1962zz}. 

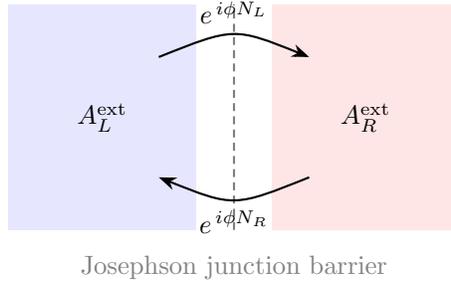
\begin{figure}[h!]
\centering
\begin{tikzpicture}[>=Stealth, every node/.style={font=\small}, thick]

  \fill[blue!10] (-3, -1.5) rectangle (-0.5, 1.5);
  \fill[red!10] (0.5, -1.5) rectangle (3, 1.5);

  \node at (-1.75, 0) {$A_L^{\mathrm{ext}}$};
  \node at (1.75, 0) {$A_R^{\mathrm{ext}}$};

  \draw[gray, thick, densely dashed] (0, -1.5) -- (0, 1.5);

  \draw[->, thick] (-1.0, 0.8) .. controls (0,1.2) .. (1.0,0.8)
      node[midway, above] {$e^{\,i\phi N_L}$};

  \draw[->, thick] (1.0,-0.8) .. controls (0,-1.2) .. (-1.0,-0.8)
      node[midway, below] {$e^{\,i\phi N_R}$};

  \node[below, gray] at (0, -1.7) {Josephson junction barrier};

\end{tikzpicture}
\caption{An algebraic depiction of the Josephson junction between two superconductors.}
\label{fig: Josephson}
\end{figure}

The Josephson effect is what happens when one takes two superconductors and brings them into interaction with each other. Algebraically, we can may naively describe the Josephson set up as a single CAR algebra $A$ in which the left superconductor defines a subalgebra $A_L \subset A$ of spinor fields supported on the left hand side of a chosen interface and similarly for the right superconductor (see Figure \ref{fig: Josephson} for a visual depiction). Equivalently, this would imply that we can think of $A_L,A_R \subset B(\mathscr{H}_\phi)$ where $\mathscr{H}_{\phi}$ is some irreducible representation of $A$. However, when two superconductors interact they pass Cooper pairs between each other. As such, the naive description of $A$ is no longer valid: there is an additional constraint which tells us that the total number of Cooper pairs must match the sum of the number of Cooper pairs on the left and the number of Cooper pairs on the right:
\beq \label{Number cons}
	N = N_L + N_R.
\eeq

To implement this constraint we must extend $A_L$ and $A_R$ to subalgebras of bounded operators on the Hilbert space $\mathscr{H}_{ext}$ described above. This extension is natural since we expect physical processes to occur which will change the relative phases of the left and right superconductors. On `global' the Hilbert space $\mathscr{H}_{ext}$ we can represent the number operators $N_L$ and $N_R$ as left and right translations
\beq
	e^{i \phi N_L} \ket{\Omega_{\phi'}} = \ket{\Omega_{\phi'+2\phi}}, \qquad e^{-i \phi N_R} \ket{\Omega_{\phi'}} = \ket{\Omega_{\phi'-2\phi}}.
\eeq
For either superconductor, the number of Cooper pairs on their side (which is equivalent to the relative number of overall Cooper pairs) becomes an observable while the number of Cooper pairs on the other side remains unobservable. In this regard, the complete set of observables for, say, the left superconductor is generated by $\pi(a)$ and $N_L$. The $C^*$ algebra generated by this pair of operators is called the $C^*$ crossed product algebra which we denote by $A_L^{ext} = A_L \times_{\alpha} U(1)$. 

The addition of the operator $N_L$ to the algebra $A_L^{ext}$ implies that it can access new physical observables. Given an overall state $\Psi \in S(A_L \times_{\alpha} U(1))$ we can now quantify transitions between vacua by computing $\Psi(e^{-i(\phi_1 - \phi_2) N_L})$. If $\ket{\Psi}$ is interpreted as a vector representative of this state in $\mathscr{H}_{ext}$ we can write
\beq
	\Psi(e^{-i(\phi_1-\phi_2)N_L}) = \bra{e^{i \phi_1 N_L} \Psi} \ket{e^{i \phi_2 N_L} \Psi},
\eeq
where the state $e^{i \phi N_L} \ket{\Psi}$ has the interpretation of creating a vacuum state of the original theory with phase $\phi$. The nonzero value of this expectation value is the mathematical manifestation of the Josephson effect -- the transitions between different `vacua' correspond with physical currents of Cooper pairs which are ferried between the two superconductors. 

\section{Perturbative Quantum Gravity} \label{sec: Gravity}

Ultimately, we would like to apply our framework towards understanding the complete background independent structure of quantum gravity. For now, that goal seems out of reach. Yet, we can still illustrate the usefulness of our framework in a more restricted sense by considering perturbative quantum gravity on null-hypersurfaces. A complimentary presentation of the following analysis geared toward the study of the generalized entropy can be found in \cite{KKFS:2025}. In the present note we wish to contextualize this analysis within the language of the framework we have described. As we will see, there is a remarkable commonality between the mathematical descriptions of null quantum gravity and the Josephson effect of superconductivity.

\subsection{Basic set up}

To formulate our gravity theory we will make use of Carrollian geometry which has been developed across a variety of works dedicated to understanding the physics of null hypersurfaces \cite{Donnay:2016ejv,Hopfmuller:2016scf,Hopfmuller:2018fni,Donnay:2019jiz,Chandrasekaran:2018aop,Adami:2021nnf,Ciambelli:2019lap,Wieland:2020gno,Freidel:2022bai,Chandrasekaran:2021vyu,Freidel:2022vjq,Freidel:2023bnj,Ciambelli:2023mir,Ciambelli:2024swv,Chandrasekaran:2023vzb}. The resulting physical picture is closely related to the infrared structure of gauge theory and gravity \cite{Sachs:1962zza,Strominger:2017zoo,Raclariu:2021zjz,Pasterski:2021rjz}, edge modes and the extended phase space \cite{Donnelly:2016auv,Ciambelli:2021nmv,Jia:2023tki,Klinger:2023qna,Geiller:2019bti}, as well as the corner symmetry paradigm \cite{Ciambelli:2021vnn,Ciambelli:2022cfr,Freidel:2020xyx,Freidel:2020svx,Freidel:2020ayo,Freidel:2021dxw,Ciambelli:2022vot}. 

A null hypersurface in spacetime can be described by a Carrollian geometry $(\mathscr{N}, q, \un{\ell}, k, D)$ where here
\begin{enumerate}
	\item $\mathscr{N}$ is a co-dimension one manifold relative to spacetime,
	\item $q$ is a degenerate metric on $\mathscr{N}$,
	\item $\un{\ell}$ is a null-normal vector on $\mathscr{N}$ satisfying $i_{\un{\ell}} q = 0$, 
	\item $k$ is a one form on $\mathscr{N}$ satisfying $i_{\un{\ell}} k = 1$ which defines a horizontal-vertical splitting\footnote{In the sequel we shall denote the horizontal projection by $\Pi: T\mathscr{N} \rightarrow H \mathscr{N}$.} of $T\mathscr{N}$ as $\un{X} = f \un{\ell} + \un{Y}$ with $i_{\un{Y}} k = 0$, and
	\item $D$ is a torsionless connection on $\mathscr{N}$ which preserves the aforementioned horizontal-vertical splitting.
\end{enumerate}
It is useful to regard $\mathscr{N}$ as fibered over a co-dimension two manifold (relative to spacetime) $\mathscr{B}$. The degenerate metric on $\mathscr{N}$ can then be pulled back to a Riemannian metric on $\mathscr{B}$ with volume form $\epsilon^{\mathscr{B}} = \sqrt{|q|} d^{D-2} x$. The manifold $\mathscr{N}$ therefore acquires a volume form $\epsilon^{\mathscr{N}} \equiv k \wedge \epsilon^{\mathscr{B}}$. 

Although the description of $\mathscr{N}$ we have given is entirely intrinsic, a natural way to obtain a Carrollian geometry is to consider the embedding of a null-hypersurface within a bulk spacetime $\mathscr{M}$. The variables $(q,\epsilon^{\mathscr{B}}, \un{\ell})$ completely characterize the bulk metric near to the hypersurface. For the purposes of our gravitational theory, these variables will play the role of `position' data in phase space. The conjugate variables to these positions are defined by the derivatives
\beq
	\bigg(\frac{1}{2} \mathcal{L}_{\un{\ell}} q\bigg)_{AB} = \frac{1}{2} \theta q_{AB} + \sigma_{AB}, \qquad D_a \epsilon^{\mathscr{N}} = -(\kappa k_a + \pi_a) \epsilon^{\mathscr{N}}. 
\eeq
Here, we have used capital Latin indices $A,B,...$ to refer to coordinates on the surface $\mathscr{B}$ and lower-case Latin indices $a,b,...$ to refer to coordinates on the full null hypersurface. The quantites $\theta$, $\sigma_{AB}$, and $\pi_a$ are, respectively, the expansion, shear and Hajicek connection.\footnote{The quantity $\kappa$ is the surface gravity.} 

A Carrollian geometry possesses an intrinsic phase space structure which is characterized by the presymplectic form\footnote{$\mu = \kappa + \frac{1}{2} \theta$.}
\beq
	\Theta^{\mathscr{N}} = \int_{\mathscr{N}} \epsilon^{\mathscr{N}} \bigg( \frac{1}{2} \sigma^{AB} \delta q_{AB} - \mu \delta(\ln \sqrt{|q|}) - \pi_a \delta \ell^a\bigg),
\eeq	
along with the pair of constraints
\begin{flalign} \label{gravitational constraints}
	&\bigg(\mathcal{L}_{\un{\ell}} + \theta\bigg) \theta = \mu \theta - \sigma_{AB} \sigma^{AB}, \nonumber \\
	&\Pi^b_a \bigg(\mathcal{L}_{\un{\ell}} + \theta\bigg) \pi_b = \Pi^b_a D_b \mu - \Pi^b_c D^c \sigma_{ab}. 
\end{flalign}
Again, there is a compelling spacetime interpretation for this structure: the presymplectic potential is equivalent to that which would have been derived from the covariant phase space upon pulling back to an embedded null hypersurface, and the constraints are equivalent to the pullback of Einstein's equations. 

\subsection{Quantization}

We will now quantize this phase space and implement its constraints perturbatively around a background spacetime admitting a bifurcate Killing horizon. We will specifically focus on the region of the resulting spacetime, $\mathscr{R}$, which is accessible to an infinitely long-lived observer. We shall denote by $\mathscr{R}'$ the complement of this region, which corresponds to the region beyond the observer's horizon. 

At linear order, the complete set of classical observables associated with the null-hypersurface is densely generated by smeared shear perturbations on the horizon
\beq
	\delta \sigma[s] \equiv \int_{\mathscr{N}} \epsilon^{\mathscr{N}} s^{AB} \delta \sigma_{AB}. 
\eeq	
By the linear Einstein equations these shears are directly related to smeared spacetime metric perturbations
\beq
	\delta g[f] = \int_{\mathscr{M}} \epsilon^{\mathscr{M}} f^{\mu \nu} \delta g_{\mu \nu} = \frac{1}{4\pi G_N} \delta \sigma[s], \qquad s = (Ef)\rvert_{\mathscr{N}}.  
\eeq
The spacetime spearing functions are required to satisfy $f^{[\mu \nu]} = 0 = \nabla_{\mu} f^{\mu \nu}$ to maintain linear diffeomorphism invariance. The object $E = G_- - G_+$ is the commutator of the linearized Einstein equations with $G_{\pm}$ the advanced/retarded propagators. 

Using the commutator we can promote the smeared metric perturbations to operators, $\op{\delta g}[f]$, governed by the commutation relations
\beq
	\bigg[\op{\delta g}[f_1], \op{\delta g}[f_2] \bigg] = i\mathbb{1} \int_{\mathscr{M}} \epsilon^{\mathscr{M}} f_1^{\mu \nu} (E f_2)_{\mu \nu}. 
\eeq
In this sense, the complete algebra of observables is nothing but the canonical commutation relation (CCR) algebra associated with linear gravity. Formally, this is the $C^*$ algebra generated by the Weyl operators $e^{i \op{\delta g}[f]}$ with algebraic relations
\beq
	e^{i \op{\delta g}[f_1]} e^{i \op{\delta g}[f_2]} = e^{-\frac{i}{2} \int_{\mathscr{M}} \epsilon^{\mathscr{M}} f_1^{\mu \nu} (E f_2)_{\mu \nu}} e^{i \op{\delta g}[f_1 + f_2]}, \qquad (e^{i \op{\delta g}[f]})^* = e^{-i \op{\delta g}[f]}.  
\eeq
In keeping with the notation of the previous section we denote this algebra by $A$. We shall denote by $A_{\mathscr{R}} \subset A$ the set of graviton operators whose smearing functions have support in $\mathscr{R}$. 

At this point in the standard approach to algebraic quantum field theory one would typically choose a quasi-free state for the algebra $A$ to provide it with a Hilbert space representation via the GNS construction.\footnote{This state is chosen to satisfy the Hadamard condition so that the short distance singularity structure of the theory is well controlled.} The GNS representation of a quasi-free state for the CCR algebra is equivalent to the standard Fock representation with the vector representative of the reference state playing the role of a vacuum. In quantum field theory on curved spacetimes there is not a unique choice of vacuum state and in fact the choice of vacuum is quite closely associated with the choice of background spacetime. Thus, we can regard the algebra $A$ as possessing a degenerate set of vacua/irreducible representations just like the CAR algebra in the BCS theory. We will revisit this point in the next section. For now, we will continue to work with the algebra $A$ abstractly, that is without specifying a particular Hilbert space representation. 

Beyond linear order, the constraints \eqref{gravitational constraints} become non-trivial. The Carrollian phase space admits a group of large diffeomorphisms which act as natural transformations and possess non-zero charges. For the full null hypersurface this group is given by
\beq
	G = \text{Diff}(\mathscr{B}) \ltimes C^{\infty}(\mathscr{B}) \ltimes C^{\infty}(\mathscr{B}),
\eeq 
which is infinitesimally generated by vector fields
\beq
	\un{X} = (f_T(x) + u f_B(x)) \frac{\partial}{\partial u} + Y^A(x) \frac{\partial}{\partial x^A}.
\eeq
Here, we are working in coordinates in which $\un{\ell} = \frac{\partial}{\partial u}$ and $x$ are coordinates for the surface $\mathscr{B}$. The vector fields $f_T(x) \frac{\partial}{\partial u}$, $u f_B(x) \frac{\partial}{\partial u}$, and $Y^A(x) \frac{\partial}{\partial x^A}$ are respectively interpreted as supertranslations, superboosts, and diffeomorphisms of the surface $\mathscr{B}$. If, additionally, we wish to preserve the region $\mathscr{R}$ we must fix the location of the surface $\mathscr{B}$ and thus we restrict to the subgroup
\beq
	G_{\mathscr{B}} \equiv \text{Diff}(\mathscr{B}) \ltimes C^{\infty}(\mathscr{B})
\eeq
generated by $\un{X} = u f_B(x) \frac{\partial}{\partial u} + Y^A(x) \frac{\partial}{\partial x^A}$. 

The group $G_{\mathscr{B}}$ acts symplectomorphically on the gravitational phase space and in turn gives rise to an automorphic action $\alpha: G_{\mathscr{B}} \rightarrow \text{Aut}(A)$. However, as was described in detail in \cite{KKFS:2025}, only the subgroup $\tilde{G}_{\mathscr{B}} \equiv G_{\textrm{isom.}} \ltimes C^{\infty}(\mathscr{B})$ preserves the GNS representation of a given Hadamard state. Here, $G_{\textrm{isom.}}$ is the group of isometries of the chosen background spacetime. If we had chosen to work within a fixed Hilbert space representation, only the subgroup $\tilde{G}_{\mathscr{B}} \subset G_{\mathscr{B}}$ would be unitarily implemented, while the remaining symmetries are spontaneously broken. Nevertheless, as was the case in the superconductivity example, we can repair the broken symmetry by gauging the full group $G_{\mathscr{B}}$ while working purely from the algebraic, rather than spatial perspective. 

To gauge the $G_{\mathscr{B}}$ symmetry we follow our approach from the analysis of the Josephson effect in superconductivity. First, we extend our global algebra to\footnote{Although $G_{\mathscr{B}}$ is not a locally compact group, we can still define a Hilbert space $L^2(G_{\mathscr{B}})$ by choosing an appropriately quasi-invariant measure. For a review of this construction, we refer the reader to Appendix \ref{App: NLC}. In \cite{KKFS:2025} we demonstrate how such a measure can be identified for the group $G_{\mathscr{B}}$ by appealing to the Euclidean gravitational path integral.} $A \otimes B(L^2(G_{\mathscr{B}}))$. This global algebra therefore obtains new operators which we quantify in terms of charges, $Q_{\phi}^{\mathscr{R}}$ and $Q_{\phi}^{\mathscr{R}'}$, which generate left and right translations on $L^2(G_{\mathscr{B}})$. In particular, if $\ket{\phi}$ is a dense set of `position' vectors for $L^2(G_{\mathscr{B}})$ we can write
\beq
	e^{i\op{Q}^{\mathscr{R}}_{\phi_1}} \ket{\phi_2} \propto \ket{\phi_1 \circ \phi_2}, \qquad e^{i \op{Q}^{\mathscr{R}'}_{\phi_1}} \propto \ket{\phi_2 \circ \phi_1^{-1}}. 
\eeq

The extended subregion algebra $A_{\mathscr{R}}^{ext}$ is then defined as the subalgebra of $A \otimes B(L^2(G_{\mathscr{B}}))$ that includes both $A$ and the algebra generated by $e^{i\op{Q}^{\mathscr{R}}_{\phi}}$ as subalgebras, and commutes with the operator $e^{i\op{Q}^{\mathscr{R}'}_{\phi}}$ for each $\phi \in G_{\mathscr{B}}$. This is precisely the $C^*$ crossed product algebra:
\beq \label{Grav CP}
	A_{\mathscr{R}}^{ext} \equiv C^*\{\pi_{\alpha}(e^{i \op{\delta g}[f]}), e^{i \op{Q}^{\mathscr{R}}_{\phi}} \; | \; \text{supp}(f) \subseteq \mathscr{R}, \phi \in G_{\mathscr{B}}\}. 
\eeq
Here,
\beq
	\pi_{\alpha}(e^{i \op{\delta g}[f]}) \equiv \alpha_{\op{\phi}^{-1}}\big( e^{i \op{\delta g}[f]} \big)
\eeq
is a graviton field dressed to the automorphism $\alpha_{\phi}$. The notation $f(\op{\phi})$ signifies a multiplication operator on the Hilbert space $L^2(G_{\mathscr{B}})$ for a sufficient nice (e.g. $L^{\infty}$) function $f \in L^{\infty}(G_{\mathscr{B}})$. That is,
\beq
	f(\op{\phi}) \ket{\phi'} = f(\phi') \ket{\phi'}. 
\eeq	
The operators $e^{i \op{Q}^{\mathscr{R}}_{\phi}}$ are unitary in the Hilbert space $L^2(G_{\mathscr{B}})$, and implement the automorphism $\alpha$ on the `dressed' gravitation operators by construction
\beq
	e^{i \op{Q}^{\mathscr{R}}_{\phi}} \pi_{\alpha}(e^{i \op{\delta g}[f]}) e^{-i \op{Q}^{\mathscr{R}}_{\phi}} = \pi_{\alpha} \circ \alpha_{\phi}(e^{i \op{\delta} g[f]}). 
\eeq

Gauging the automorphism $\alpha$ can be understood as implementing the charge-flux relation implied by Einstein's equations. Each vector field $\un{X} \in \mathfrak{g}_{\mathscr{B}} \equiv \text{Lie}(G_{\mathscr{B}})$ can be associated with a classical phase space observable $F_{\un{X}}$ which implements the associated (infinitesimal) diffeomorphism through the Poisson bracket as
\beq \label{Classical derivation}
	\{F_{\un{X}},\delta g[f]\} = \delta g[\mathcal{L}_{\un{X}} f]. 
\eeq
Using the gravitational constraints \eqref{gravitational constraints} we can write 
\beq
	F_{\un{X}} = \int_{\mathscr{N}} d q_{\un{X}},
\eeq
with $q_{\un{X}}$ is a $D-2$ form on $\mathscr{N}$. Invoking Stokes' theorem we find that
\beq \label{Charge-Flux}
	F_{\un{X}} = Q^{\mathscr{R}}_{\un{X}} - Q^{\mathscr{R}'}_{\un{X}},
\eeq
where $Q^{\mathscr{R}/\mathscr{R}'}_{\un{X}}$ are the co-dimension two integral of $q_{\un{X}}$ on the boundaries of $\mathscr{N}$ which are in $\mathscr{R}/\mathscr{R}'$. The charges $Q^{\mathscr{R}/\mathscr{R}'}_{\un{X}}$ depend directly upon higher order perturbations of the background and therefore cannot belong to the algebra $A$. As we have addressed, in the quantum theory it is not immediately clear how to identify $F_{\un{X}}$ with a self-adjoint operator. Nevertheless, we can still regard \eqref{Classical derivation} as a derivation of the gravitational algebra which exponentiates to the automorphism $\alpha$. Interpreting the constraint \eqref{Charge-Flux} in this light, it can be understood at the level of automorphic maps as\footnote{Here, we have identified $\un{X}$ with the diffeomorphism $\phi$ that it generates.}
\beq \label{Charge-Flux Aut}
	\alpha_{\phi} = \text{Ad}_{\op{Q}^{\mathscr{R}}_{\phi}} \circ \text{Ad}_{\op{Q}^{\mathscr{R}'}_{\phi}}^{-1}.
\eeq	
This condition is satisfied by the dressed algebra \eqref{Grav CP} in which operators commute with $\op{Q}_{\phi}^{\mathscr{R}'}$ and the automorphism $\alpha_{\phi}$ is implemented by $\op{Q}_{\phi}^{\mathscr{R}}$. 

\begin{figure}[h!]
\centering
\begin{tikzpicture}[>=Stealth, thick, every node/.style={font=\small}]

  \coordinate (O) at (0,0);
  \coordinate (L1) at (-2.5,-2.5);
  \coordinate (L2) at (-2.5,2.5);
  \coordinate (R1) at (2.5,-2.5);
  \coordinate (R2) at (2.5,2.5);

  \draw[thick, dashed, gray] (O) -- (2.5,2.5);
  \draw[thick, dashed, gray] (O) -- (2.5,-2.5);

  \fill[blue!10] (-2.5,-2.5) -- (O) -- (-2.5,2.5) -- cycle;
  \node[blue!60!black] at (-1.3,0) {$A^{\mathrm{ext}}_{\mathscr{R}}$};

  \fill[red!10] (O) -- (2.5,-2.5) -- (2.5,2.5) -- cycle;
  \node[red!60!black] at (1.3,0) {$A^{\mathrm{ext}}_{\mathscr{R}'}$};

  \node at (-1.2,2.2) {$\mathscr{R}$};
  \node at (1.2,2.2) {$\mathscr{R}'$};

  \draw[->, thick] (-0.6,0.7) .. controls (0,1.2) .. (0.6,0.7)
      node[midway, above] {$e^{\,i\op{Q}^{\mathscr{R}}_{\phi}}$};

  \draw[->, thick] (0.6,-0.7) .. controls (0,-1.2) .. (-0.6,-0.7)
      node[midway, below] {$e^{\,i\op{Q}^{\mathscr{R}'}_{\phi}}$};

  \node[black] at (O) {$\mathscr{B}$};

  \draw[->, thin, gray!70] (0,-2.8) -- (0,3) node[above] {$t$};
  \draw[->, thin, gray!70] (-3,0) -- (3,0) node[right] {$x$};

\end{tikzpicture}
\caption{The constraint physics of a horizon creates a gravitational version of a Josephson junction.}
\label{fig: gravitational-junction}
\end{figure}
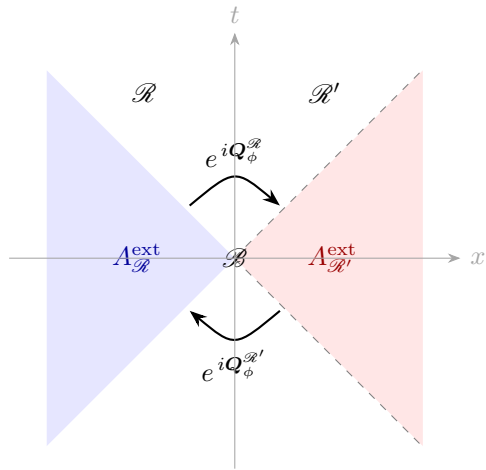

Eqn. \eqref{Charge-Flux Aut} constitutes a global constraint which should be compared to \eqref{Number cons}. In this regard, we see that gauging the automorphism $\alpha$ creates a situation which may be thought of as a gravitational version of a Josephson junction. This is depicted graphically in Figure \ref{fig: gravitational-junction}.

\subsection{Vacuum Interpretation}

A useful way to interpret the algebra $A_{\mathscr{R}}^{ext}$ is to regard it as a concrete algebra of observables acting on the Hilbert space $\mathscr{H}_{ext} \equiv \int_{G_{\mathscr{B}}}^{\oplus} d\nu(\phi) \mathscr{H}_{\phi}$. Here, $\nu$ is a quasi-invariant measure on $G_{\mathscr{B}}$ and each $\mathscr{H}_{\phi}$ is the GNS Hilbert space obtained from the vacuum state associated with the spacetime $(\mathscr{M},\phi^*g)$ with $g$ the metric we perturbed around in our analysis.

Following the notation set forward in Section \ref{sec: BCS}, we can introduce a family of vacuum states $\ket{\Omega_{\phi}}$ labeling the different spacetime geometries which can be obtained by acting on the background with a diffeomorphism contained in $G_{\mathscr{B}}$. Likewise, we can denote by $\pi_{\phi}: A \rightarrow B(\mathscr{H}_{\phi})$ the family of associated GNS representations. A general operator can be written in the form
\beq
	f \equiv \int_{G_{\mathscr{B}}}^{\oplus} d\nu(\phi) \pi_{\phi}(e^{i \op{\delta g}[f(\phi)]}),
\eeq	 
where $f(\phi)$ is a graviton smearing function for each $\phi \in G_{\mathscr{B}}$. This operator creates states
\beq
	\ket{f} \equiv \int_{G_{\mathscr{B}}}^{\oplus} d\nu(\phi) \pi_{\phi}(e^{i \op{\delta g}[f(\phi)]}) \ket{\Omega_{\phi}} = \int_{G_{\mathscr{B}}}^{\oplus} d\nu(\phi) \ket{f(\phi)}_{\phi},
\eeq
with each $\ket{f(\phi)}_{\phi}$ interpreted as a graviton coherent state with smearing $f(\phi)$ on top of the background $(\mathscr{M},\phi^*g)$.  

At the same time, this Hilbert space also admits the translation operator
\beq
	e^{i \op{Q}^{\mathscr{R}}_{\phi_1}} \ket{\Omega_{\phi_2}} \equiv \ket{\Omega_{\phi_1 \circ \phi_2}}.
\eeq
This operator should be interpreted as an intertwiner mapping between geometries $\phi_2^* g$ and $(\phi_1 \circ \phi_2)^* g$. Defining the representation $\pi: A \rightarrow B(\mathscr{H}_{ext})$ by
\beq
	\pi(e^{i\op{\delta g}[f]}) \equiv \pi_{\phi}(e^{i\op{\delta g}[f]}),
\eeq
we see that the intertwining operators $e^{i\op{Q}^{\mathscr{R}}_{\phi}}$ generate the automorphism $\alpha$ unitarily. Thus, by \cite{DoplicherKastlerRobinson1966}, the $C^*$ algebra generated by $\pi(e^{i\op{\delta g}[f]})$ and $e^{i \op{Q}^{\mathscr{R}}_{\phi}}$ is isomorphic to the crossed product $A_{\mathscr{R}}^{ext}$. 

In summary, we can think of the algebra $A_{\mathscr{R}}^{ext}$ and the Hilbert space $\mathscr{H}_{ext}$ as being defined across two dimensions -- (QFT) vacua and excitations. This is visualized in figure \ref{fig: Perturbative gravity}. For a fixed (QFT) vacuum, the operators $\pi(e^{i\op{\delta g}[f]})$ create graviton excitations. The operators $e^{i \op{Q}^{\mathscr{R}}_{\phi}}$ generate transitions between different geometries which intertwine between (QFT) vacuum states.

\begin{figure}[h!]
\centering
\begin{tikzpicture}[>=Stealth, every node/.style={font=\small}]

  \draw[->, thick] (0,0) -- (5,0) node[right] {QFT Vacua/Backgrounds};
  \draw[->, thick] (0,0) -- (0,5) node[above] {Excitations};

  \coordinate (A) at (1.5,2.0);
  \coordinate (B) at (2.8,2.0);
  \coordinate (C) at (2.8,3.2);

  \fill (A) circle (2pt);
  \fill (B) circle (2pt);
  \fill (C) circle (2pt);

  \draw[->, thick] (A) -- (B) node[midway, above] {$e^{\,i\op{Q}^{\mathscr{R}}_{\phi}}$};
  \draw[->, thick] (B) -- (C) node[midway, right] {$\displaystyle \pi\big(e^{\,i\op{\delta g}[f]}\big)$};

  \draw[dotted] (A) -- (A |- 0,0);
  \draw[dotted] (B) -- (B |- 0,0);
  \draw[dotted] (C) -- (C |- 0,0);

  \draw[dotted] (A) -- (A -| 0,0);
  \draw[dotted] (B) -- (B -| 0,0);
  \draw[dotted] (C) -- (C -| 0,0);

  \node[below] at (A |- 0,0) {$\phi'$};
  \node[below] at (B |- 0,0) {$\phi \circ \phi'$};
  \node[left] at (A -| 0,0) {$f'$};
  \node[left] at (C -| 0,0) {$f + f'$};

\end{tikzpicture}
\caption{Visualization of the Hilbert space $\mathscr{H}_{\mathrm{ext}}$ and the algebra $A^{\mathrm{ext}}_{\mathscr{R}}$.}
\label{fig: Perturbative gravity}
\end{figure}

\section{General Framework, Applications and Future Directions} \label{sec: Discussion}

The algebra $A^{ext}_{\mathscr{R}}$ affords a much richer picture than conventional local QFT lore. We have accommodated many states which otherwise would have been regarded as superselected in local QFT by introducing new physical operators that map between these states. These operators quantify geometric fluctuations and seem to be closely related to vacuum transition amplitudes. We are hopeful that the algebra $A^{ext}_{\mathscr{R}}$ may therefore be useful in exploring recent proposals for how to construct observables in quantum gravity that are accessible in realistic experimental set ups \cite{He:2024vlp,Badurina:2024rpp,Ciambelli:2025flo,Fransen:2025npa,He:2025hag}. We expect that the geometric fluctuations of perturbative quantum gravity should prove to be an important phenomenological feature of the theory, just as the Josephson effect has become for superconductivity \cite{osti_4644689}. 

With this being said, the present analysis still only allows for transitions between geometries that belong to the same `family'. That is, we can only intertwine spacetimes which are related to an initially chosen `overall' background $(\mathscr{M},g)$ by a diffeomorphism belonging to the group $G_{\mathscr{B}}$. As we have alluded to in the introduction, this seems to bear a close resemblance to the observation of \cite{Liu:2025cml,Liu:2025krl} that concrete algebras in string theory are referenced to a class of asymptotic geometries. Nevertheless, it is our hope that the framework outlined in this note can provide a new perspective on the study of backgrounds for quantum systems that lends itself towards understanding even non-perturbative quantum gravity where transitions between these `overall backgrounds' are also allowed. There are obviously many questions that still need to be answered before we can get to that point. Still, we would now like to provide a rather broad overview as to how the approach of the present note may figure into answering some of these questions.

\subsection{The General Framework} \label{sec: framework}

One of the most important conceptual themes of our framework is that we should view `backgrounds' as distinct representations for a particular abstract system. In our language the abstract system is encoded in a $C^*$ algebra $A$ and the representations correspond to different Hilbert spaces on which $A$ is realized concretely as an operator algebra. Through the GNS construction, the specification of unitarily inequivalent representations for the algebra $A$ is closely related to the problem of defining abstract states on $A$ and determining their domains. From a mathematical point of view, these facts seem to suggest that the correct framework for studying backgrounds is the collection of (equivalence classes of) unitarily inequivalent representations of $A$ which we denote by $\text{Rep}(A)$. It is typically not necessary to consider the complete collection $\text{Rep}(A)$. Rather, we often restrict our attention to a subclass of physically allowable representations $\text{Rep}_a(A) \subset \text{Rep}(A)$. For example, in local quantum field theory we only admit GNS representations associated with Hadamard states. 

Under some circumstances the collection of allowable representations obtains the structure of a category.\footnote{It is not necessary to have a great familiarity with category theory to follow this discussion. Briefly, one can think of the category of allowed representations as a generalization of a group where different representations are thought of as charges, and the structure of the category tells us how to combine these charges.}  In \cite{AliAhmad:2025bnd}, we explained how to derive from\footnote{Our construction is valid for the case that $\text{Rep}_a(A)$ is a multi-fusion category. This roughly means that it has a finite set of charges and admits well defined rules for combining charges. It is possible we will need a generalization of this construction in quantum gravity to allow for `infinite' representation categories. See the discussion section of \cite{AliAhmad:2025bnd} for some comments to this end.} $\text{Rep}_a(A)$ a generalized symmetry acting on the algebra $A$. Formally, this follows from the Tannaka-Krein construction which tells us that every fusion category is functorially equivalent to the representation category of a weak Hopf algebra. A weak Hopf algebra, also called a quantum groupoid, is a far reaching generalization of a group. As a quantum group can be obtained from a regular group by deforming its algebraic relations, a quantum groupoid can be obtained from a groupoid. 

To understand the spirit of the Tannaka-Krein result, let's consider the simplest case where $\text{Rep}_a(A)$ is equivalent to the representation category of a group $G$. The statement $\text{Rep}_a(A) \simeq \text{Rep}(G)$ tells us that the inequivalent representations of $A$ can be identified with irreducible unitary representations of the group $G$. This is precisely what we observed in the superconductivity example where the degenerate vacua of the BCS algebra were labeled by $U(1)$ charges. In this regard, we could have derived the spontaneously broken $U(1)$ symmetry on the BCS algebra simply by studying its physically allowed representations. To move beyond the fixed background regime in superconductivity we studied interacting subsystems which could pass Cooper pairs between each other. The algebra of the superconducting subsystem was thereby extended to a crossed product $A_L \times_{\alpha} G$ (and similarly for the right) in which the newly added operators precisely map between previously superselected vacuum states. We note that irreducible representations of $A_L \times_{\alpha} G$ are naturally identified with reducible representations of $A_L$.  

In the general case, the correspondence $\text{Rep}_a(A) \simeq \text{Rep}(H)$ for $H$ a weak Hopf algebra is simply a relabeling of the different allowable backgrounds for the system $A$. It tells us that different vacua of the system $A$ can be identified with `charges' of the symmetry $H$.\footnote{This is very closely related to the sense in which different backgrounds in string theory can be identified with VEVs of dynamical fields in the complete background independent theory. We discuss this point further below.} To achieve background independence we can again prepare a pair of interacting subsystems and thereby pass to the generalized crossed product $A_L^{ext} \equiv A_L \times_{\alpha} H$. Here, $\alpha$ is the generalized symmetry action of the weak Hopf algebra $H$ on the system $A$. We refer the reader to \cite{AliAhmad:2025bnd} for a detailed construction of this action and its associated crossed product. Schematically, $A_L^{ext}$ is simply generated by dressed operators satisfying a charge conservation constraint and new operators called `charged intertwiners' that map between inequivalent representations of $A$. We note in passing that this form of gauging bears a very sharp resemblance to the notion of SymTFT entanglement introduced in \cite{Torres:2025jcb}. There, as here, a generalized symmetry is gauged by gluing together two theories that share a common symmetry subcategory. In \cite{Torres:2025jcb}, this is accomplished by utilizing the idea of SymTFT, in which a $d$-dimensional theory and its generalized symmetries are unified into a single $d+1$-dimensional topological theory. In \cite{Evans:2025msy}, the SymTFT was given an algebraic avatar in terms of the theory of subfactor inclusions of $C^*$ algebras. Appealing to \cite{AliAhmad:2025oli,AliAhmad:2025bnd}, such inclusions have an interpretation as our generalized crossed products. We hope to address the connection between these ideas more closely in future work.

The preceding discussion appears to suggest the following procedure for constructing background independent algebras. Start by encoding the system of interest in terms of an abstract $C^*$ algebra, $A$. Find the set of allowed Hilbert space representations of $A$, $\text{Rep}_a(A)$. These correspond to the different backgrounds relative to which observables in $A$ can be measured. Using Tannaka-Krein or a generalization, the set of allowed representations of $A$ can be identified with charges of a quantum symmetry algebra $H$. This algebra can be seen to act on $A$ as a generalized symmetry. If $A$ is placed into a bath and allowed to exchange charges with its environment the set of observable degrees of freedom is enlarged to the algebra $A \times_{\alpha} H$ which includes new intertwining operators mapping between what were superselected vacuum states relative to $A$. This can be viewed as an approach to gauging the generalized symmetry on $A$ in accord with the principle that quantum gravity should admit no global symmetries. The algebra $A \times_{\alpha} H$ is background independent in the sense that one can move freely between any background in the course of measuring observables.

The above procedure presents an analytic framework for addressing the question of what the algebra of observables is for a closed universe in quantum gravity, and the closely related question of the `observer'. The correspondence between backgrounds, representations, and generalized symmetries emphasizes the fact that the lack of global symmetries in quantum gravity is a manifestation of background independence. Even after performing our generalized gauging we run into the following problem: If the representations of the algebra of interest correspond to the set of backgrounds (and equivalently quantify the global symmetries of the system) then the only truly background independent algebra is the one with a trivial category of representations. One may view this as an argument in favor of the one-dimensional Hilbert space for a closed universe. However, there also appears to be a natural resolution -- any `actual' measurement that can be made is relational and therefore at least partially breaks background independence. Put differently, a particular `observer' is only allowed to measure a subclass of charges. Thus, it would only be necessary to gauge a subset of \emph{admissible} representations to obtain the physical algebra of observables for such an observer. This point of view provides a nice formalization of the notion that an observer can be described by a code subspace of states \cite{Witten:2023xze}, or, put differently, a partial gauge fixing. The role of the observer is  quite analogous to the way that phases of a system can be described by their broken symmetries. The non-trivial algebra accessible to the observer is characterized by a non-trivial category of representations which itself corresponds to a broken symmetry that the observer implicitly references their measurements to.  

There are several aspects of this construction which require additional input both physically and mathematically. The first question is what algebra $A$ we should start with. A good candidate 
would be the algebra described in the locally covariant approach to local quantum field theory \cite{Brunetti:2006qj,Fredenhagen:2011hm,Brunetti:2013maa,Brunetti:2022itx}. Assuming an algebra $A$ can be constructed, the next problem is determining its allowable representations. From there, we can hope to place a categorical structure on this collection to translate it into a generalized symmetry. This will probably require a more broad notion of symmetry category than what is presently understood in the literature, and by consequence a more powerful form of Tannaka-Krein duality. Finally, we will need to understand how these more general symmetries act on the algebra $A$ and the algebraic extensions associated with them.

It is useful at this point to compare again with the discussion of \cite{Liu:2025cml,Liu:2025krl}. There, it is argued that, in the algebraic language, string field theory defines a concrete algebra of observables $M_{\omega}$ for each choice of background state $\omega$ corresponding to a different class of asymptotic boundary conditions on spacetime. More generally, a choice of background state in string theory may be identified with the asymptotic value of any dynamical field in the theory \cite{McNamara:2020uza}. We can interpret each $M_{\omega} \equiv \pi_{\omega}(A_{\textrm{IIB}})''$ as defining a different concrete representation of some underlying, background independent algebra $A_{\textrm{IIB}}$ which is yet unknown. The above discussion suggests an approach to reconstructing the complete quantum gravity algebra from the data contained in the collection of such representations. This can be viewed in many respects as a vast generalization of the reconstruction theorem proven by Doplicher and Roberts (DR) \cite{Doplicher:1989}.\footnote{In DR reconstruction, the reconstructed algebra is viewed as kinematical rather than physical. Its usefulness is that it is generated by unobservable, but more computationally useful field operators while the physical net is generated only by gauge invariant objects. In the gravity context, the reconstructed algebra is gauge invariant since the intertwining operator are, in principle, observable. With this being said, from the point of view of a `realistic' observer who lives in a spacetime with fixed asymptotic data we can still move from the reconstructed algebra back to the `observer's algebra' by fixing a representation.} The expectation that the complete non-perturbative string theory admits all of the representations $M_{\omega}$ simultaneously (e.g. that backgrounds arise from VEVs of dynamical fields in the complete theory) can be viewed as an illustration of this generalized DR reconstruction.

At least at present, it is not attainable to fully reconstruct the algebra $A_{\textrm{IIB}}$ or a similarly general algebra. With this being said, it does seem possible that such a reconstruction could be carried out for a more simple e.g. topological string theory. This may be related to recent work \cite{Barbar:2023ncl,Dymarsky:2024frx,Barbar:2025vvf}, which constructed a holographic duality between $3d$ gravity and an ensemble of CFTs by starting with a general $3d$ TQFT and gauging its maximal non-anomalous symmetry group. This results in a `gravitational' TQFT in which the gauged symmetry gives rise to a sum over equivalence classes of topologies. 

In any case, the framework of studying gravitational algebras via DR like reconstruction can help to clarify many puzzles that have been circulating in recent literature. As an illustration, in subsection \ref{App: BU}, we describe how recent work toward understanding holography for closed universes may be related to this point of view. Especially, the notion of subsystem in quantum gravity may center around determining which representations of the (perhaps unattainable) background independent algebra are allowed for a given observer. In other words, a subsystem in quantum gravity is a subcategory of the complete category of representations which may be DR reconstructed to a subalgebra of the overall background independent algebra. In principle, this can be accomplished without detailed knowledge of the complete algebra or its full representation category. We are hopeful that this point of view might clarify the physical origin of the `observer' that has appeared in the construction of gravitational operator algebras, and in organizing a consistent quantum mechanical interpretation of the gravitational path integral. 

\subsection{Baby Universes as `Quantum Symmetries'} \label{App: BU}

To conclude, we'd like to make some brief remarks about how the framework we have initiated can be related to recent work on holography for closed universes \cite{Liu:2025cml,Gesteau:2025obm,Kudler-Flam:2025cki}. This discussion follows closely the general set up described in \cite{Liu:2025cml}.\footnote{We are grateful to Hong Liu for insightful discussion about this work.} The identification of baby universes as a form of quantum symmetry bears a resemblance to the study of $\alpha$-parameters in the seminal work of Coleman \cite{Coleman:1988cy}, Giddings and Strominger \cite{Giddings:1988wv}, and later Marolf and Maxfield \cite{Marolf:2020xie}. These $\alpha$-parameters label different superselection sectors of an effective gravitational theory in which the contribution of baby universes has been integrated out. From an algebraic point of view, this suggests that the complete gravity algebra $A_{ext}$ should be realized as a direct integral of effective algebras for each $\alpha$, along with the explicit baby-universe creation and annihilation operators which move between different $\alpha$-sectors. This is an example of the general archetype of algebra we have advocated for in subsection \ref{sec: framework}. We will now discuss how a related structure can be gleaned when viewing holography from the algebraic point of view, although we leave to future work the explicit connection (if it exists) with the standard $\alpha$-parameters.  

Let $\mathscr{H}_{\Psi} \equiv \{\mathscr{H}_N, \ket{\Psi_N}\}_{N \in \mathbb{N}}$ denote a family of quantum theories specified by Hilbert spaces $\mathscr{H}_N$ along with preferred states $\ket{\Psi_N} \in \mathscr{H}_N$. In the context of the AdS/CFT correspondence, we regard $\mathscr{H}_{\Psi}$ as encoding boundary conformal field theories with $N$ indexing the rank of the gauge group. We assume that there is a well defined limiting Hilbert space $\mathscr{H}_{\Psi} \simeq \lim_{N \rightarrow \infty} \mathscr{H}_N$. Here, we have used the same notation to denote the limiting Hilbert space as we have to denote the sequence that defines it. The Hilbert space $\mathscr{H}_{\Psi}$ obtains a preferred state $\ket{\Psi} \in \mathscr{H}_{\Psi}$ which is defined as the limit of the family of states $\Psi \equiv \{\ket{\Psi_N} \in \mathscr{H}_N\}_{N \in \mathbb{N}}$.

In the AdS/CFT correspondence, the state $\ket{\Psi}$ is referred to as a semiclassical state. This is because it is dual to a solution to the bulk gravity equations of motion which we denote by $g_{\Psi}$. On the gravity side, we can define a Hilbert space $\mathscr{H}_{g_{\Psi}}^{F}$ which is the Fock space of small excitations quantized around this solution. The AdS/CFT correspondence can subsequently be understood as an isomorphism between the two Hilbert spaces we have thus far constructed:
\beq \label{AdS/CFT HS}
	\mathscr{H}_{\Psi} \simeq \mathscr{H}_{g_{\Psi}}^{F}. 
\eeq
Let us emphasize that different Hilbert spaces $\mathscr{H}_{\Psi}$ can be obtained by considering different sequences of CFTs, different preferred states on these CFTs, or both. On the gravity side these choices would lead to different semiclassical backgrounds via \eqref{AdS/CFT HS}. 

An operator $A \in B(\mathscr{H}_{\Psi})$ is said to have a well defined large-$N$ limit \emph{in the state} $\ket{\Psi}$ if it can be identified with a family of operators $A \equiv \{A_N \in B(\mathscr{H}_N)\}_{N \in \mathbb{N}}$ such that $\lim_{N \rightarrow \infty} \bra{\Psi_N} A_N \ket{\Psi_N} < \infty$. We assume that the set of operators with well defined large-$N$ limits with respect to $\Psi$ form an algebra which we denote by $\mathcal{A}_{\Psi}$. The vector $\ket{\Psi}$ defines an algebraic state on $\mathcal{A}_{\Psi}$ as $\omega_{\Psi}(A) \equiv \bra{\Psi} A \ket{\Psi}$. By construction, the set of large $N$-operators with respect to $\Psi$ form a dense domain for the state $\omega_{\Psi}$ and thus we can also interpret the Hilbert space $\mathscr{H}_{\Psi}$ as the GNS Hilbert space of the algebra $\mathcal{A}_{\Psi}$ with respect to $\omega_{\Psi}$.  We shall denote by $\pi_{\Psi}: \mathcal{A}_{\Psi} \rightarrow B(\mathscr{H}_{\Psi})$ the associated GNS representation. The weak closure of $\mathcal{A}_{\Psi}$ in the topology induced by the representation $\pi_{\Psi}$ is a von Neumann algebra $\mathcal{X}_{\Psi} \equiv \pi_{\Psi}(\mathcal{A}_{\Psi})''$. If $\omega_{\Psi}$ is a pure state on $\mathcal{A}_{\Psi}$, which is typically assumed to be true, the representation $\pi_{\Psi}$ will be irreducible and we can identify $\mathcal{X}_{\Psi} \simeq B(\mathscr{H}_{\Psi})$. 

To understand the physics of this set up and relate it to our framework, it is useful to recast our analysis around an abstract algebra together with its collection of representations. In the AdS/CFT context, a good candidate arises in the form of the (large $N$ limit of the) algebra of single trace operators in the CFT. In contrast to the algebra $\mathcal{A}_{\Psi}$ which is defined with respect to a chosen state $\Psi$, the algebra of single trace operators is \emph{universal} in the sense that it should obtain a well defined large-$N$ limit with respect to \emph{any} semiclassical state. Let us refer to this algebra by $\mathcal{S}$. Explicitly, $\mathcal{S} \equiv \lim_{N \rightarrow \infty} \mathcal{S}_N$, where $\mathcal{S}_N \subset B(\mathscr{H}_N)$ is the collection\footnote{At finite $N$ the collection of single trace operators doesn't close in an algebra. Nonetheless, it is still useful to consider this set and its limiting properties.} of single trace operators in the CFT at each finite $N$. From this point of view, a semiclassical state $\Psi$ defines a concrete representation of the otherwise abstractly defined single trace algebra, $\pi_{\Psi}: \mathcal{S} \rightarrow B(\mathscr{H}_{\Psi})$. 

In the context of subregion-subalgebra duality, the algebra $\mathcal{Y}_{\Psi} \equiv \pi_{\Psi}(\mathcal{S})'' \subseteq \mathcal{X}_{\Psi}$ is dual to the causal wedge of the full boundary theory \cite{Leutheusser:2022bgi}. Each representation $\pi_{\Psi}$ therefore also identifies an inclusion $\mathcal{Y}_{\Psi} \subseteq \mathcal{X}_{\Psi}$. If $\omega_{\Psi}$ is a pure state when restricted to $\mathcal{S}$, we will find that $\mathcal{Y}_{\Psi} \simeq \mathcal{X}_{\Psi} \simeq B(\mathscr{H}_{\Psi})$. Conversely, if $\omega_{\Psi}$ is a mixed state when restricted to $\mathcal{S}$, we will find that $\mathcal{Y}_{\Psi} \subset \mathcal{X}_{\Psi}$ is a strict inclusion. The latter case implies that the causal wedge is entangled with a non-trivial algebra of operators $\mathcal{H}_{\Psi} \equiv \mathcal{Y}_{\Psi}' \wedge \mathcal{X}_{\Psi}$. In \cite{Liu:2025cml}, it is argued that the algebra $\mathcal{H}_{\Psi}$ should be interpreted as an algebra of baby universe operators which act on a sector of the theory that is beyond the causal wedge. These different scenarios can be depicted in a compelling way in the bulk -- as seen in figure (2) of \cite{Liu:2025cml}. When $\omega_{\Psi}\rvert_{\mathcal{S}}$ is pure, the complete bulk algebra is described by a two-sided black hole featuring two copies of AdS which are disconnected besides the standard entanglement of the thermofield double state. When $\omega_{\Psi}\rvert_{\mathcal{S}}$ is mixed, the complete bulk algebra is described by two copies of AdS that are glued together along a baby universe. This baby universe facilitates further entanglement between the copies by allowing for operators to intertwine between the two geometries. This picture should be compared to Figures \ref{fig: Josephson} and \ref{fig: gravitational-junction}.

At this stage, we should emphasize the no-go theorem proposed by Gesteau in \cite{Gesteau:2025obm} which claims that if the semiclassical state $\Psi$ exists, then it must be pure on the single trace algebra. In \cite{Liu:2025cml,Kudler-Flam:2025cki} it has been argued that this theorem may be circumvented by modifying how the large $N$ limit is taken. The purpose of the present discussion is to explore the algebraic consequences \emph{under the assumption} that a limit resulting in a mixed state does indeed exist. In particular, we'd like to argue that the resulting  structure reproduces the kind of algebraic extension which naturally emerges from the framework outlined in subsection \ref{sec: framework} above. Going in the opposite direction, then, it is tempting to conjecture that the problem of understanding holography for closed universe (e.g. making sense of a large $N$ limit which \emph{can} accommodate a mixed semiclassical state) can be translated into an analysis of the backgrounds/background independence of the abstract single trace algebra. We will comment further on this claim shortly.

Building upon the above discussion, let us consider the following set up on the boundary side. At each finite $N$ the Hilbert space is of the form $\mathscr{H}_N \equiv \mathscr{H}_N^{(L)} \otimes \mathscr{H}_N^{(R)}$, where $\mathscr{H}_{N}^{(L)}/\mathscr{H}_{N}^{(R)}$ will be dual to the left/right AdS geometries above in the appropriate limit. Let $\mathcal{S}^{(L)}/\mathcal{S}^{(R)}$ denote the abstract algebras associated with single trace operators in the left and right CFTs. Likewise, let $\mathcal{X}_{\Psi}^{(L)}/\mathcal{X}_{\Psi}^{(R)}$ denote the complete algebras of large $N$ operators associated with the left and right CFTs with respect to the semiclassical state $\Psi$. In relation to the notation used above, we have $\mathcal{S} = \mathcal{S}^{(L)} \vee \mathcal{S}^{(R)}$ and $\mathcal{X}_{\Psi} = \mathcal{X}_{\Psi}^{(L)} \vee \mathcal{X}_{\Psi}^{(R)}$. 

Given a semiclassical state $\Psi \equiv \{\ket{\Psi_N}\}_{N \in \mathbb{N}}$, we can consider the following modular flows. Firstly, we have the modular flow $\sigma^{\Psi \mid \mathcal{X}_{\Psi}^{(R)}}: \mathbb{R} \rightarrow \text{Aut}(\mathcal{X}_{\Psi}^{(R)})$, which arises from the state $\omega_{\Psi}$ restricted to the full right CFT algebra. At the same time, for each $N$ we have the modular flow $\sigma^{\Psi_N \mid \mathcal{S}_N^{(R)}}: \mathbb{R} \rightarrow \text{Aut}(\mathcal{S}_N^{(R)})$, which arises from the state $\omega_{\Psi_N}$ restricted to the collection of right single trace operators. Suppose that $s \equiv \lim_{N \rightarrow \infty} s_N$ is a (right) single trace operator obtained from a limit of finite $N$ single trace operators $\{s_N\}_{N \in \mathbb{N}}$. In \cite{Leutheusser:2022bgi}, it is shown that
\beq \label{Liu-Leutheusser}
	\pi_{\Psi}\bigg(\lim_{N \rightarrow \infty} \sigma^{\Psi_N \mid \mathcal{S}_N^{(R)}}_t(s_N) \bigg) = \sigma^{\Psi \mid \mathcal{X}_{\Psi}^{(R)}}_t \circ \pi_{\Psi}(s). 
\eeq
It is tempting to conjecture that the left-hand side of \eqref{Falcone-Takesaki} can be rewritten in the form
\beq \label{Falcone-Takesaki}
	 \pi_{\Psi}\bigg(\lim_{N \rightarrow \infty} \sigma^{\Psi_N \mid \mathcal{S}_N^{(R)}}_t(s_N) \bigg) = \pi_{\Psi} \circ \sigma^{\Psi \mid \mathcal{S}^{(R)}}_t(s),
\eeq
where $\sigma^{\Psi \mid \mathcal{S}^{(R)}}: \mathbb{R} \rightarrow \text{Aut}(\mathcal{S}^{(R)})$ is the modular automorphism of the state $\omega_{\Psi}$ restricted to the algebra of (right) single trace operators. Eqn. \eqref{Falcone-Takesaki} simply distributes the limit to both the modular flow ($\sigma^{\Psi_N \mid \mathcal{S}_N^{(R)}}_t \rightarrow \sigma^{\Psi \mid \mathcal{S}^{(R)}}_t$) and the operator being flowed ($s_N \rightarrow s$). The limit is made subtle, however, by the fact that the single trace operators do not close an algebra at each finite $N$. 

If eqn. \eqref{Falcone-Takesaki} is valid, we obtain the equality
\beq \label{Falcone-Takesaki 2}
	\pi_{\Psi} \circ \sigma^{\Psi \mid \mathcal{S}^{(R)}}_t(s) = \sigma^{\Psi \mid \mathcal{X}_{\Psi}^{(R)}}_t \circ \pi_{\Psi}(s), \qquad \forall s \in \mathcal{S}^{(R)}. 
\eeq
A theorem by Falcone and Takesaki \cite{FalconeOVW} states that \eqref{Falcone-Takesaki 2} implies the existence of an operator valued weight $T_{\Psi}: \mathcal{X}_{\Psi}^{(R)} \rightarrow \mathcal{Y}_{\Psi}^{(R)}$ such that $\omega_{\Psi}\rvert_{\mathcal{X}_{\Psi}^{(R)}} = \omega_{\Psi}\rvert_{\mathcal{S}^{(R)}} \circ T_{\Psi}$. Consequently, we can regard $\mathcal{Y}_{\Psi}^{(R)} \subset \mathcal{X}_{\Psi}^{(R)}$ as an algebraic extension in the sense described in \cite{AliAhmad:2025oli}. From this point of view, the dual of the baby universe algebra $\mathcal{H}_{\Psi}^*$ is naturally interpreted as a quantum symmetry algebra admitting an action $\rho: \mathcal{H}_{\Psi}^* \rightarrow \text{End}(\mathcal{Y}_{\Psi}^{(R)})$ such that $\mathcal{X}_{\Psi}^{(R)} \simeq \mathcal{Y}_{\Psi}^{(R)} \times_{\rho} \mathcal{H}_{\Psi}^*$.\footnote{We refer the reader to Section 4 of \cite{AliAhmad:2025oli} for an explanation of the notation $\mathcal{Y}_{\Psi}^{(R)} \times_{\rho} \mathcal{H}_{\Psi}^*$ in terms of spatial Q-systems.} The algebra $\mathcal{X}_{\Psi}$ likewise acquires an induced action of $\mathcal{H}_{\Psi}$ for which $\mathcal{Y}_{\Psi}^{(R)}$ is the invariant subalgebra. This reproduces the observation that $\mathcal{H}_{\Psi} \simeq \mathcal{Y}_{\Psi}^{(R)}{}' \wedge \mathcal{X}_{\Psi}$. If we moreover assume that $\omega_{\Psi}$ remains a normalized state before and after restriction to single trace operators this map is actually a conditional expectation. In this case, it may be possible to further refine the classification of $\mathcal{H}_{\Psi}$ in accord with the discussion in Appendix E of \cite{AliAhmad:2025bnd}.\footnote{This suggests that the collection of baby universe operators should form a weak Hopf algebra; this conclusion has some interesting synergy with the observations of \cite{Gesteau:2024gzf}.}

Following the arguments of \cite{Leutheusser:2022bgi}, it is natural to expect that the inclusion $\mathcal{Y}_{\Psi} \subset \mathcal{X}_{\Psi}$ is a boundary manifestation of the inclusion of the causal wedge within the `entanglement wedge'.\footnote{The geometric region probed in our analysis extends the usual entanglement wedge and should probably be given a more representative name.} Under more typical holographic assumptions, it has been claimed that we should not expect \eqref{Falcone-Takesaki 2} to hold. Rather, it has been argued that the full algebra $\mathcal{X}_{\Psi}$ is generated by modular flow on the single trace algebra $\mathcal{Y}_{\Psi}$. This is regarded as a statement of entanglement wedge reconstruction. One way of interpreting the present analysis is that the same modification to the large $N$ limit which allows for the existence of semiclassical mixed states on the single trace algebra will also lead to a modification of the limit appearing in \eqref{Falcone-Takesaki}. Because this large $N$ limit probes a sector which was invisible in the standard large $N$ theory, there is an obstruction to the modular flow of the single trace algebra generating all of $\mathcal{X}_{\Psi}$. This obstruction results in the emergence of the generalized symmetry algebra which we have identified with the baby universe. 

In this light, choosing the state $\Psi$ or equivalently the operator valued weight $T_{\Psi}$ constitutes a generalization of the usual quantum extremal surface prescription and should be governed by an entropic minimization protocol. Crucially, this version of bulk reconstruction is fully algebraic and fully boundary anchored. A similar proposal was made in \cite{AliAhmad:2024saq} with regards to the emergence of geometry from purely algebraic considerations. From the point of view described in the main text, this extremization program coincides with determining the set of allowed representations for $\mathcal{S}^{(R)}$. In this regard, the constraints on the optimization problem should be interpreted as specifying a particular `observer'. We plan to address the problem of characterizing $\mathcal{H}_{\Psi}$ in relation to the theory of inclusions and to present an associated algebraic extremization prescription in future work. 

As we have mentioned, such a prescription would in effect \emph{define} a modified large $N$ limit which can accommodate semiclassical mixed states. Indeed, the algebraic structure described above appears to be consistent with the mathematical notion of an ultrafilter.\footnote{Ultrafilters and their relation to large $N$ limits were recently discussed in the interesting work \cite{Chen:2025uzn}. We are grateful to Tom Faulkner for introducing us to the notion of ultrafilters and their usefulness in understanding the large-$N$ limit in holography.} An ultrafilter of the natural numbers is a collection of subsets of $\mathbb{N}$ which does not contain the empty set, is closed under intersections and supersets, and contains either a set or its complement but not both. Given a sequence, $(a_n)_n$, of elements in some topological space $A$, an ultrafilter, $\Omega$, can be used to define a generalized limit:
\beq
	\lim_{n \rightarrow \Omega} a_n = a \iff \{n \in \mathbb{N} \; | \; a_n \in U_a \; \forall \text{ neighborhoods of }a\} \in \Omega.
\eeq
In the case that the ultrafilter is chosen to contain cofinite subsets (e.g. sets whose complements are finite) this limit agrees with the usual $\lim_{n \rightarrow \infty}$. We will denote the standard cofinite ultrafilter by $\Omega_{cf}$. Using the machinery of ultrafilters, Ocneanu developed a rigorous notion of the limit of a sequence of operator algebras \cite{Ocneanu1985}. That is, given a sequence of von Neumann algebras with preferred states, $(M_n,\varphi_n)_n$, Ocneanu's procedure outputs a von Neumann algebra $M_{\Omega}$. This algebra naturally admits an inclusion $M_{\Omega_{cf}} \subset M_{\Omega}$, where $M_{\Omega_{cf}}$ is the algebra of operators possessing standard pointwise limits, and this inclusion admits a canonical (at least relative to the choice $\Omega$) conditional expectation $E_{\Omega}: M_{\Omega} \rightarrow M_{\Omega_{cf}}$ \cite{ANDO20146842}. The relative commutant of the inclusion is non-trivial, $H_{\Omega} \equiv M_{\Omega_{cf}}' \wedge M_{\Omega}$, and defines what is called the Golodet asymptotic algebra \cite{Golodets1975}. Heuristically, we may regard this as the algebra of operators which obtain a limit through the ultrafilter $\Omega$, but otherwise would not converge pointwise in $n$. 

The analogs between this construction and the above discussion are very enticing. We would like to identify $\mathcal{X}_{\Psi} \sim M_{\Omega}$, $\mathcal{Y}_{\Psi} \sim M_{\Omega_{cf}}$, and $\mathcal{H}_{\Psi} \sim H_{\Omega}$. However, the ultrafilter comes with its own set of mathematical question marks. Most dramatically, the limiting algebra depends on the choice of ultrafilter, and may inherit certain pathologies from the limiting process. Nevertheless, it is encouraging that this notion of generalized large-$N$ limit brings about the same inclusion structure we have identified. In this regard, it again seems reasonable to expect that there may exist a physical criterion by which a preferred ultrafilter (and by extension a preferred large $N$ limit) is identified. We look forward to exploring these ideas further in future work.

\appendix
\renewcommand{\theequation}{\thesection.\arabic{equation}}
\setcounter{equation}{0}

\section*{Acknowledgments}
We thank Shadi Ali Ahmad, Luca Ciambelli, Tom Faulkner, Elliott Gesteau, Temple He, Jonah Kudler-Flam, Gabriele La Nave, Albert Law, Rob Leigh, Hong Liu, Daniel Murphy, Gautam Satishchandran, Antony Speranza, Michael Stone, Yifan Wang, and Kathryn Zurek for many valuable discussions. This work was supported by the Heising-Simons foundation ``Observable Signatures of Quantum Gravity" collaboration and the Walter Burke Institute for Theoretical Physics. This material is also based upon work supported by the U.S. Department of Energy, Office of Science, Office of High Energy Physics, under Award Number DE-SC0011632 

\section{$C^*$ Algebras and Canonical Purifications} \label{app: GNS}

A $C^*$ algebra is a Banach space $A$ equipped with three compatible operations:
\begin{enumerate}
	\item Multiplication:\footnote{Multiplication must be associative: $(ab) c = a (bc)$.} $a,b \in A \mapsto ab \in A$,
	\item Involution:\footnote{The involution must be involutive: $(a^*)^* = a$ and $(ab)^* = b^* a^*$.} $a \in A \mapsto a^* \in A$,
	\item Norm: $a \in A \mapsto \norm{a} \in \mathbb{C}$. 
\end{enumerate}
The compatibility between these three operations can be encoded in the $C^*$ relation
\beq
	\norm{a^* a} = \norm{a}^2. 
\eeq
A state on a $C^*$ algebra is a positive, linear, normalized map $\omega: A \rightarrow \mathbb{C}$. One can think of the assignment $a \in A \mapsto \omega(a) \in \mathbb{C}$ as the computation of an expectation value. We shall denote the space of states on $A$ as $S(A)$. A state $\omega \in S(A)$ is called pure with respect to the algebra $A$ if and only if $\omega = p \omega_1 + (1-p) \omega_2$ implies that $\omega = \omega_1 = \omega_2$. Otherwise the state is called mixed. 

The abstract notions of algebras and states described above can be brought into a more familiar form by appealing to the notion of a representation. A Hilbert space representation of $A$ is a map $\pi: A \rightarrow B(\mathscr{H})$ which is compatible with the defining operations of the algebra as
\beq
	\pi(ab) = \pi(a) \pi(b), \qquad \pi(a^*) = \pi(a)^{\dagger}, \qquad \norm{\pi(a)}_{B(\mathscr{H}} = \norm{a}. 
\eeq
Here, $\dagger$ is the natural involution on $B(\mathscr{H})$ which is induced from the inner product on $\mathscr{H}$ e.g.
\beq
	\langle \xi_1, \pi(a) \xi_2 \rangle_{\mathscr{H}} = \langle \pi(a)^{\dagger} \xi_1, \xi_2 \rangle_{\mathscr{H}},
\eeq
and $\norm{\cdot}_{B(\mathscr{H}}$ is the standard operator norm.  A representation $\pi$ is called a purification for a state $\omega \in S(A)$ if there exists a vector representative $\xi_{\omega} \in \mathscr{H}$ such that
\beq
	\omega(a) = \langle \xi_{\omega}, \pi(a) \xi_{\omega} \rangle_{\mathscr{H}}. 
\eeq	
As the nomenclature suggests, the state $\langle \xi_{\omega}, \cdot \xi_{\omega} \rangle_{\mathscr{H}}$ is pure on the full algebra $B(\mathscr{H})$, though it may not be pure when restricted to $\pi(A) \subset B(\mathscr{H})$. 

Given any state $\omega \in S(A)$ there exists a canonical purification which can be obtained through the Gelfand-Naimark-Segal (GNS) construction. We first define the domain of the state by $\mathfrak{n}_{\omega} \equiv \{a \in A \; | \; \omega(a^*a) < \infty\}$ and the kernel of the state by $\mathfrak{k}_{\omega} \equiv \{a \in A \: | \; \omega(a^* a) = 0\}$. Let $\ket{a}_{\omega}$ denote the equivalence class of the operator $a$ modulo addition by elements in the kernel of $\omega$. That is $\ket{a}_{\omega}$ belongs to the quotient space $\mathfrak{n}_{\omega}/\mathfrak{k}_{\omega}$. This quotient space is a pre-closed inner product space with pre-inner product
\beq
	{}_{\omega}\bra{a_1} \ket{a_2}_{\omega} \equiv \omega(a_1^* a_2). 
\eeq
The completion of $\mathfrak{n}_{\omega}/\mathfrak{k}_{\omega}$ with respect to this inner product defines a Hilbert space which we denote by $\mathscr{H}_{\omega}$. 

On this Hilbert space we have a natural representation $\pi_{\omega}: A \rightarrow B(\mathscr{H}_{\omega})$ given by left multiplication
\beq
	\pi_{\omega}(a_1) \ket{a_2}_{\omega} \equiv \ket{a_1 a_2}_{\omega}. 
\eeq
There is an obvious cyclic state in this Hilbert space which can be represented by $\ket{\mathbb{1}}_{\omega}$. One may interpret $\ket{\mathbb{1}}_{\omega}$ as the `vacuum' state, with all the other states in the Hilbert space obtained by exciting this state with operators $a \in A$, e.g. $\ket{a}_{\omega} = \pi_{\omega}(a) \ket{\mathbb{1}}_{\omega}$. We also see that
\beq
	{}_{\omega}\bra{\mathbb{1}} \pi_{\omega}(a) \ket{\mathbb{1}}_{\omega} = {}_{\omega}\bra{\mathbb{1}} \ket{a}_{\omega} = \omega(a). 
\eeq	
Thus, the representation $\pi_{\omega}$ is a purification for the state $\omega$ with vector representative $\ket{\mathbb{1}}_{\omega}$. 

\section{The Group Algebra for Non-Locally Compact Groups} \label{App: NLC}

A measure $\nu$ on a group $G$ is termed quasi-invariant if there exists a dense subgroup $G_{\nu} \subseteq G$ and a cocycle $J_{\nu}: G_{\nu} \times G \rightarrow \mathbb{R}$ such that
\beq
	d\nu(gh) = J_{\nu}(g,h) d\nu(h), \qquad J_{\nu}(gh,k) = J_{\nu}(g,hk)J_{\nu}(h,k), \qquad g,h \in G_{\nu}, \; k \in G. 
\eeq
Given such a quasi-invariant measure we may define a Hilbert space $L^2(G,d\nu)$ consisting of square integrable functions on $G$ with respect to $\nu$. This Hilbert space admits a natural unitary representation $\ell_{\nu}: G_{\nu} \rightarrow U(L^2(G,d\nu))$ given by
\beq
	\bigg(\ell_{\nu}(g) \xi\bigg)(h) \equiv J_{\nu}(g,g^{-1}h)^{1/2} \xi(g^{-1} h). 
\eeq
The group algebra of $G$ with respect to $\nu$ is the $C^*$ algebra generated by this unitary representation.

\providecommand{\href}[2]{#2}\begingroup\raggedright\endgroup

\end{document}